\documentclass[preprintnumbers, floatfix,letterpaper,aps,prd,epsfig,nofootinbib,
onecolumn
]{revtex4-1}
\pdfoutput=1
\usepackage{bm,graphicx,dcolumn,epstopdf,epsf, latexsym,mathbbol, amssymb,amsmath,color,slashed, mathrsfs,mathcomp,simplewick}
\usepackage[center]{subfigure}
\usepackage{multirow}
\usepackage{makecell}
\usepackage[colorlinks,linkcolor=blue,citecolor=blue,urlcolor=blue]{hyperref}

\begin{document}
\allowdisplaybreaks
 \newcommand{\bq}{\begin{equation}}
 \newcommand{\eq}{\end{equation}}
 \newcommand{\bqn}{\begin{eqnarray}}
 \newcommand{\eqn}{\end{eqnarray}}
 \newcommand{\nb}{\nonumber}
 \newcommand{\lb}{\label}
 \newcommand{\f}{\frac}
 \newcommand{\p}{\partial}
\newcommand{\PRL}{Phys. Rev. Lett.}
\newcommand{\PLB}{Phys. Lett. B}
\newcommand{\PRD}{Phys. Rev. D}
\newcommand{\CQG}{Class. Quantum Grav.}
\newcommand{\JCAP}{J. Cosmol. Astropart. Phys.}
\newcommand{\JHEP}{J. High. Energy. Phys.}
\newcommand{\bea}{\begin{eqnarray}}
\newcommand{\ena}{\end{eqnarray}}
\newcommand{\beqa}{\begin{eqnarray}}
\newcommand{\eeqa}{\end{eqnarray}}
\newcommand{\red}{\textcolor{black}}

\title{Waveform of gravitational waves in the general parity-violating gravities}

\author{Wen Zhao${}^{a, b}$}
\email{Corresponding author: wzhao7@ustc.edu.cn}

\author{Tao Zhu${}^{c, d}$}

\author{Jin Qiao${}^{c, d}$}

 \author{Anzhong Wang${}^{e}$}

\affiliation{${}^{a}$ CAS Key Laboratory for Research in Galaxies and Cosmology, Department of Astronomy, University of Science and Technology of China, Hefei 230026, China; \\
${}^{b}$ School of Astronomy and Space Sciences, University of Science and Technology of China, Hefei, 230026, China\\
${}^{c}$ Institute for theoretical physics and cosmology, Zhejiang University of Technology, Hangzhou, 310032, China\\
${}^d$ United Center for Gravitational Wave Physics (UCGWP), Zhejiang University of Technology, Hangzhou, 310032, China\\
${}^{e}$ GCAP-CASPER, Physics Department, Baylor University, Waco, TX 76798-7316, USA}

\date{\today}

%
%

\begin{abstract}
As an extension of our previous work \cite{qiao}, in this article, \red{we calculate the effects of parity violation on}
gravitational-wave (GW) waveforms \red{during their propagation}
in the most general parity-violating gravities, including Chern-Simons modified gravity, ghost-free scalar-tensor gravity, symmetric teleparallel equivalence of GR theory, Ho\v{r}ava-Lifshitz gravity and so on. \red{For this purpose, we consider the GWs generated by the coalescence of compact binaries and concentrate on the imprints of the parity violation in the propagation of GWs.} With a unified description of GW in the theories of parity-violating gravity, we study the effects of velocity and amplitude birefringence on the GW waveforms. Decomposing the GWs into the circular polarization modes, the two birefringence effects exactly correspond to the modifications in phase and amplitude of GW waveforms respectively. We find that, for each circular polarization mode, the amplitude, phase and velocity of GW can be modified by both the parity-violating terms and parity-conserving terms in gravity. Therefore, in order to test the parity symmetry in gravity, we should compare the difference between two circular polarization modes, rather than measuring an individual mode. Combining two circular modes, we obtain the GW waveforms in the Fourier domain, and obtain the deviations from those in General Relativity. The GW waveforms derived in this paper are also applicable to the theories of parity-conserving gravity, which have the modified dispersion relations (e.g. massive gravity, double special relativity theory, extra-dimensional theories, etc), or/and have the modified friction terms (e.g. nonlocal gravity, gravitational theory with time-dependent Planck mass, etc).
\end{abstract}


\maketitle

\section{Introduction}
\renewcommand{\theequation}{1.\arabic{equation}} \setcounter{equation}{0}

Gravitational waves (GWs) are always produced in the circumstances with extreme conditions (e.g. the strongest gravitational field, the densest celestial bodies, the earliest stage of the universe, the highest energy scale physics, etc), and have the weak interactions with other matters during the propagation \cite{pn,spa,maggiore2}. Therefore, the GWs encode the cleanest information for these extreme conditions, and provides the excellent opportunity to study the physics in these extreme circumstances. As an example of the applications, GWs can be used to test the theory of gravity \cite{yunes-nature,sathya-white}, which has become an important topic in the era of GW astronomy \cite{gw150914,gw170817,gw-other,gw150914-testGR,gw170817-testGR,gw170817-speed,lorentz2, lrr}. Although Einstein's General Relativity (GR) has been considered to be the most successful theory of gravity since it was proposed, it faces the difficulties in both theoretically (e.g. singularity, quantization, etc), and observationally (e.g. dark matter, dark energy, etc). Therefore, testing GR in various circumstance is an important topic since its birth \cite{test1,test2,test3,test4,test5,test6}.

As well known, symmetry permeates nature and is fundamental to all laws of physics. Thus, one important method for the gravity examination is to test the symmetries in gravity. In our series of works, we focus on the testing of parity symmetry in gravity. Parity symmetry implies that a directional flipping to the left and right does not change the laws of physics. It is well known that nature is parity violating. Since the first discovery of parity violation in weak interactions \cite{Lee-Yang}, the experimental tests become more necessary in the other interactions, including gravity. Although, the parity symmetry is maintained in GR, the gravitational terms with parity violation are always motivated by anomaly cancelation in particle physics and string theory \cite{CS-review,string}, and various parity-violating (PV) theories of gravity haven been proposed in literature, including the Chern-Simons modified gravity \cite{JackiwPi,cs1,cs2,cs3,cs4}, ghost-free scalar-tensor gravity \cite{ghost1,ghost2,gao}, symmetric teleparallel equivalence of GR theory \cite{tele}, Ho\v{r}ava-Lifshitz gravity \cite{Horava:2009uw,soda,soda2,Zhu:2011xe,wang_polarizing_2012,zhu_effects_2013,Wang:2017brl} and so on. As an extension of previous work \cite{qiao}, in this article, we will calculate the waveform of GWs, produced by the coalescence of compact binaries, in the general framework of PV gravities.

The effect of parity violation on GW is the birefringence phenomenon during its propagation, i.e. the symmetry between the left-hand and right-hand circular polarization modes is broken \cite{CS-review,ghost2,qiao,zhao2019}. For instance, the primordial GWs generated in the PV gravities are circularly polarized, which leave the significant imprints in the temperature and polarization anisotropies of cosmic microwave background radiation \cite{lue,soda,wang_polarizing_2012,zhu_effects_2013,Alexander:2004wk,smolin}.
However, it seems no hope to detect these imprints in the near future \cite{nohope}, due to the small amplitude of primordial GWs. As one of our series of works, in this article, we will study the imprints of parity violation in gravity on the waveforms of GW, produced by the coalescence of compact binaries. Decomposing the GWs into the circular polarization modes, we find their propagation equations are decoupled, and the deviations from those in GR, caused by the PV terms, are quantified by two parameterized terms $\nu_{A}$ and $\mu_A$, which exactly correspond to the effects of velocity and amplitude birefringence respectively. As an application of these results, we focus on four specific theories of PV gravity (Chern-Simons gravity, ghost-free scalar-tensor gravity, symmetric teleparallel equivalence of GR theory, Ho\v{r}ava-Lifshitz gravity), and obtain the explicit expressions of these two terms for each theory. Converting the circular polarization modes to the general plus and cross modes, for the general theories of PV gravity, we obtain the GW waveforms in the Fourier domain. Due to the mixture of plus and cross modes during their propagation, we find both velocity and amplitude birefringence effects can contribute the modifications in GW phase and amplitude, and the deviations of GW waveforms from those in GR are also derived.

This paper is organized as follows. In Sec. \ref{sec2}, in the effective field theory of gravity, we study the field equation of GW in the general theories of PV gravity. In Sec. \ref{sec3}, we focus on the specific models of PV gravity, and obtain the expression of corresponding quantities, which violate the parity symmetry. In Sec. \ref{sec4}, we discuss the amplitude and velocity birefringence effects of GWs. In Sec. \ref{sec5}, we calculate the waveform of GWs produced by the coalescence of compact binary systems, and particularly focus on the deviations from those in GR. The summary of this work is given in Sec. \ref{sec6}.

Throughout this paper, the metric convention is chosen as $(-,+,+,+)$, and greek indices $(\mu,\nu,\cdot\cdot\cdot)$ run over
$0,1,2,3$ \red{and latin indices $(i, \; j,\;k)$ run over $1, 2, 3$}. We set the units to $c=\hbar=1$.

\section{Gravitational waves in the general parity-violating gravities\label{sec2}}
\renewcommand{\theequation}{2.\arabic{equation}} \setcounter{equation}{0}

We consider the general theories of gravity with parity violation, whose action can be written as follows,
\bea\label{first-action}
S=\frac{1}{16\pi G} \int {d}^4 x \sqrt{-g} [\mathcal{L}_{\rm GR}+\mathcal{L}_{\rm PV}+\mathcal{L}_{\rm other}],
\ena
where $\mathcal{L}_{\rm GR}$ is the Einstein-Hilbert term $R$ or its equivalent quantity in the non-Riemannian formulation
of gravity \cite{tele}. $\mathcal{L}_{\rm PV}$ represents the PV Lagrangian density, which always consist of a number of terms in the specific models. $\mathcal{L}_{\rm other}$ is the Lagrangian density for the other matters, the scalar field and the modification terms of gravity, which are not relevant to parity violation.


In the flat Friedmann-Robertson-Walker (FRW) universe, GW is the tensor perturbation of the metric, {{i.e.,}}
\begin{equation}
ds^2=a^2(\tau)\left[-d\tau^2+(\delta_{ij}+h_{ij})dx^i dx^j\right],
\end{equation}
where $a(\tau)$ is the conformal scale factor, and we set its present value as $a_0=1$ in this paper. $\tau$ is the conformal time, which relates to the cosmic time by $dt=ad\tau$. $x^i$ is the comving coordinates. The quantity $h_{ij}$ stands for the GW perturbation, which we take to be transverse and traceless gauge, $\delta^{ij}h_{ij}=0$ and $\partial_i h^{ij}=0$.

The equation of motion of GW is determined by the tensor quadratic action, which reads \cite{grishchuk,weinberg,mtw}
\bea
S^{(2)}=\frac{1}{16\pi G} \int dt d^3x~ a^3\left[\mathcal{L}^{(2)}_{\rm GR}+\mathcal{L}^{(2)}_{\rm PV}+\mathcal{L}^{(2)}_{\rm other}\right],
\ena
where
\bea
\mathcal{L}^{(2)}_{\rm GR}=\frac{1}{4}\left[\dot{h}_{ij}^2-a^{-2}(\partial_{k} h_{ij})^2\right],
\ena
is the standard Lagrangian obtained from the Einstein-Hilbert term $R$. In this article, a {\emph{dot}} denotes the derivative with respect to
the cosmic time $t$. In the viewpoint of effective fields \cite{effective-field}, the operators $\dot{h}^2_{ij}$ and $(\partial_{k}h_{ij})^2$ are the only quadratic operators with two derivatives. Note that, in principle this Lagrangian can be generalized by adding a time-dependent coefficient in each term, which might be caused by the Horndeski or Galileon scalar field, through the perturbation of extrinsic curvature of the spatial slices \cite{horn}. Similar terms also appear in the nonlocal gravity \cite{nonlocal}, Ho\v{r}ava-Lifsthitz gravity \cite{wang_polarizing_2012} and so on. Therefore, in this paper, {we consider the extended version of the above Lagrangian, which takes the form},
\bea
\mathcal{L}^{(2)}_{\rm eGR}=\frac{1}{4}\left[b_1(t)\dot{h}_{ij}^2-b_2(t)a^{-2}(\partial_{k} h_{ij})^2\right],
\ena
where $b_1$ and $b_2$ are the arbitrary functions of time. {In order to avoid ghost instability, one has to require $b_1(t)>0$.} The propagation effects of GWs, due to these modifications, have been analyzed explicitly in the previous works \cite{arai-123,ezquiaga}.

The first possible corrections to the tensor mode mentioned above come from terms with three derivatives. There are only two possible terms \cite{effective-field},
\bea
\epsilon^{ijk}\dot{h}_{il}\partial_{j}\dot{h}_{kl},~~\epsilon^{ijk}\partial^2 {h}_{il}\partial_{j}{h}_{kl},
\ena
where $\epsilon^{ijk}$ is the antisymmetric symbol. So, the standard quadratic action is modified by the addition of \cite{effective-field}
\bqn\label{gw-action}
\mathcal{L}^{(2)}_{\rm PV}=\frac{1}{4} \left[\frac{c_1}{a \red{M_{\rm PV}} }\epsilon^{ijk} \dot{h}_{il}\partial_{j}\dot{h}_{kl}+\frac{c_2}{a^3\red{M_{\rm PV}}}\epsilon^{ijk}\partial^2 h_{il}\partial_{j} h_{kl}\right],
\eqn
where a {\emph{dot}} denotes the derivative with respect to the cosmic time $t$, $c_1$ and $c_2$ are dimensionless coefficients, and $\red{M_{\rm PV}}$ is the scale that suppresses these higher dimension operators. \red{Normally, $M_{\rm PV}$ can be constrained by the solar system experiments and various astrophysical observations. Since most of the experimental and astrophysical constraints are considered in the Chern-Simons (CS) modified gravity, here we consider constraints in  this theory as examples to discuss the possible order of the magnitude of $M_{\rm PV}$. In the solar system experiment, the frame-dragging measurement with LAGEOS places the bound $M_{\rm PV}^{-1} \lesssim 2000 {\rm km}$ \cite{smith}. For astrophysical test, the binary pulsar observations yield $M_{\rm PV}^{-1} \lesssim 0.4 {\rm km }$ \cite{yunes}. The possible bounds on $M_{\rm PV}$ have also been explored with the detection of the GWs, and as pointed out in \cite{cs4, zhao2019},  the observation of the GWs from the compact binaries are expected to provide more strong bound on $M_{\rm PV}$.} In this paper, we assume the homogeneous and isotropic background of the universe, and take $c_1$ and $c_2$ to be functions of time.

The equation of motion for GWs can be derived by varying the quadratic action with respect to $h_{ij}$. We consider the GWs propagating in the vacuum, and the field equation for $h_{ij}$ is given by,
\bqn\label{aaaa}
b_1h_{ij}'' + (2 \mathcal{H}b_1+b_1') h_{ij}'  - b_2\partial^2 h_{ij} + \frac{\epsilon^{ilk}}{a\red{M_{\rm PV}}} \partial_l \Big[ c_1 h_{jk}'' + (\mathcal{H}c_1+c_1') h_{jk}' - c_2 \partial^2 h_{jk}\Big]=0,
\eqn
where a {\emph{prime}} denotes the derivative with respect to the conformal time $\tau$, and $\mathcal{H}\equiv a'/a$. \red{When $b_1=1=b_2$, the above equation reduces to the same form as Eq.~(3.9) in \cite{qiao} in the ghost-free parity violating gravities. This is easy to understand because the GR is only modified by the parity violating terms in the action of the ghost-free parity violating gravities and $c_1$, $c_2$ represent their effects on the propagation equation of GWs \cite{qiao}. However, in the general case, as one can see from (\ref{first-action}), the action of the theory can also contain other matter terms or modifications of gravity, which are labeled by $\mathcal{L}_{\rm other}$ in (\ref{first-action}) and are not relevant to parity violation. $\mathcal{L}_{\rm other}$ can also modify the propagation equation of GWs and their effects are directly characterized by the parameters $b_1$ and $b_2$ in (\ref{aaaa}). }

In the PV gravities, it is convenient to decompose the GWs into the circular polarization modes. To study the evolution of $h_{ij}$, we expand it over spatial Fourier harmonics,
\bqn
h_{ij}(\tau, x^i) = \sum_{A={\rm R, L}} \int \frac{d^3 k}{(2\pi)^3}  h_{\rm A}(\tau, k^i)e^{i k_i x^i} e_{ij}^{\rm A}(k^i),
\eqn
where $e_{ij}^{\rm A}$ denotes the circular polarization tensors and satisfy the relation {$\epsilon^{i j k} n_j e_{kl}^A = i \rho_A e^{i A}_{~l}$}, with $\rho_{\rm R}=1$ and $\rho_{\rm L} =-1$. So, the equation of motion in Eq.(\ref{aaaa}) can be written as \cite{ghost2,zhao2019,qiao}
{
\bqn
h_{\rm A}''+\left(2+\frac{b_1'/\mathcal{H} + \rho_A (k/a\red{M_{\rm PV}})(c_1-c_1'/\mathcal{H})}{b_1- \rho_A (k/a\red{M_{\rm PV}}) c_1}\right)\mathcal{H}h'_{\rm A}+\left(1+\frac{b_2-b_1 +\rho_A (k/a\red{M_{\rm PV}})(c_1-c_2)}{b_1- \rho_A (k/a\red{M_{\rm PV}}) c_1}\right)k^2 h_{\rm A}=0.
\eqn
We expect the deviations from GR is small such that the terms with $\mathcal{O}(b_1'),  \mathcal{O}(c_1),  \mathcal{O}(c_1'),  \mathcal{O}(c_2),  \mathcal{O}(b_2-b_1) \ll 1$. With this consideration, the above equation can be further simplified into the form, }
\bea\label{A8}
h_{\rm A}''+(2+\bar{\nu}+\nu_{\rm A})\mathcal{H}h'_{\rm A}+(1+\bar{\mu}+\mu_{\rm A})k^2 h_{\rm A}=0,
\ena
where ${\rm A=R }$ or ${\rm L}$, standing for the right-hand or left-hand polarization mode respectively, and
\bea\label{b1b2c1c2}
{\mathcal{H} \bar{\nu}= (\ln{b_1})',~~}
\bar{\mu}=b_2/b_1-1,~~
{\mathcal{H} \nu_{\rm A}=- \big[\rho_{\rm A}(k/a\red{M_{\rm PV}}) (c_1/b_1) \big]', ~~}
\mu_{\rm A}=\rho_{\rm A}(k/a\red{M_{\rm PV}})(c_1-c_2)/b_1. \label{A8b}
\ena
\red{The new effects arising from the generic PV gravities are fully characterized by four parameters: $\bar \nu$, $\bar \mu$, $\nu_A$, and $\mu_A$. While the parameters $\nu_A$ and $\mu_A$ label the effects of the parity violation, the parameters $\bar \nu$ and $\bar \mu$ arise from the other possible modifications which are not relevant to parity violation. In these four parameters, $\bar \mu$, $\mu_A$ determine the speed of the gravitational waves and $\bar \nu$, $\nu_A$ provide an amplitude modulation to the gravitational waveform. Specific to parity violation, the parameter  $\mu_A$ leads to different velocities of left-hand and right-hand circular polarizations of GWs, therefore the arrival times of the two circular polarization modes could be different. For parameter $\nu_A$, it is easy to see that it has the same value but opposite signs for the left-hand and right-hand GWs. As a result, the amplitude of left-hand circular polarization of gravitational waves will increase (or decrease) during the propagation, while the amplitude for the right-hand modes will decrease (or increase). }

As an extension of the above theory, we can generalize the expressions of $\bar\nu$, $\bar\mu$, $\nu_{\rm A}$ and $\mu_{\rm A}$ as the following parameterized forms
\bea\label{coes}
{\mathcal{H}\bar{\nu}= [\alpha_{\bar{\nu}}(\tau)(k/a\red{M_{\rm PV}})^{\beta_{\bar{\nu}}}]',~~}
\bar{\mu}=\alpha_{\bar{\mu}}(\tau)(k/a\red{M_{\rm PV}})^{\beta_{\bar{\mu}}},~~
{\mathcal{H}\nu_{\rm A}= [\rho_{\rm A}\alpha_{\nu}(\tau)(k/a\red{M_{\rm PV}})^{\beta_{\nu}}]',~~}
\mu_{\rm A}=\rho_{\rm A}\alpha_{\mu}(\tau)(k/a\red{M_{\rm PV}})^{\beta_{\mu}},\nb\\
\ena
where $\beta_{\bar{\nu}}$ and $\beta_{\bar{\mu}}$ are the arbitrary numbers, and $\beta_{{\nu}}$ and $\beta_{{\mu}}$ are the arbitrary odd numbers. $\alpha_{\bar{\nu}}$, $\alpha_{\bar{\mu}}$, $\alpha_{\nu}$, $\alpha_{\mu}$ are the arbitrary functions of time. The formula in Eq.(\ref{A8}) with coefficients in Eq.(\ref{coes}) is the unifying description for low-energy effective description of generic PV GWs. To our knowledge, all the known theories of PV gravity in the literature, even if including more than three-derivative terms, can be casted into this form, which will be presented explicitly in Sec. \ref{sec3}. Here, we should stress that, even for the theories of gravity with parity symmetry, the equation of motion of GWs is also described by this formula in (\ref{A8}). For instance, in the specific case with the only nonzero quantity $\bar{\mu}$ and $\beta_{\bar\mu}=-2$, this formula becomes the motion of equation of GW in the massive gravity \cite{will}.

\section{Specific models of alternative gravities \label{sec3}}
\renewcommand{\theequation}{3.\arabic{equation}} \setcounter{equation}{0}

\subsection{Chern-Simons modified gravity}

We first consider the Chern-Simons (CS) modified gravity, which has been widely studied in the previous works \cite{JackiwPi,cs1,cs2,cs3,cs4}. CS modified gravity is an effective extension of GR that captures leading-order, gravitational PV term. The similar versions of this theory were suggested in the context of string theory \cite{string}, and three-dimensional topological massive gravity \cite{massive}. In this theory, the action in gravity is given by Eq.(\ref{first-action}) and the PV term is \bqn\label{cs-action}
\mathcal{L}_{\rm PV} = \frac{1}{8}\vartheta(\phi) \varepsilon^{\mu\nu\rho\sigma} R_{\rho\sigma \alpha\beta} R^{\alpha \beta}_{\;\;\;\; \mu\nu},
\eqn
with $\varepsilon_{\rho \sigma \alpha \beta}$ being the Levi-Civit\'{a} tensor defined in terms of the the antisymmetric symbol $\epsilon^{\rho \sigma \alpha \beta}$ as $\varepsilon^{\rho \sigma \alpha \beta}=\epsilon^{\rho \sigma \alpha \beta}/\sqrt{-g}$. The quantity $\vartheta$ is the so-called CS coupling field. If $\vartheta=$const. CS modified gravity reduces to identically to GR, since the Pontryagin term in Eq.(\ref{cs-action}) can be expressed as the divergence of the CS topological current \cite{CS-review}. Therefore, in the dynamical CS gravity, $\vartheta$ is an arbitrary function of the scalar field $\phi$, which is a function of spacetime. Similar to the previous works \cite{cs1,cs2,cs3,cs4}, in the FRW universe, we assume that the scalar field $\phi$ is a function of conformal time $\tau$ only to ensure that the background symmetry is preserved. The equation of motion of GW is determined by the tensor quadratic action, which in the CS modified gravity is given by Eq.(\ref{gw-action}) where the coefficients are $c_1 = c_2 = \dot{\vartheta}\red{M_{\rm PV}}$. Therefore, the propagation equation of GW becomes that in (\ref{A8}) with
\bqn
{\mathcal{H} \bar \nu = 0,~~}
\bar{\mu}=0,~~
{\mathcal{H}\nu_{\rm A} = - \big[ \rho_A c_1 (k/a\red{M_{\rm PV}}) \big] ', ~~ }
\mu_{\rm A}=0.
\eqn
The corresponding values of the parameters ($\alpha_{\bar{\nu}}$,$\beta_{\bar{\nu}}$,$\alpha_{\bar{\mu}}$,$\beta_{\bar{\mu}}$,$\alpha_{\nu}$,$\beta_{\nu}$,$\alpha_{\mu}$,$\beta_{\mu}$) defined in Eq.(\ref{coes}) are listed in Table {\ref{tab1}}.
Note that, CS modified theory has higher-derivative field equations, which induces the dangerous Ostrogradsky ghosts \cite{CS-review}. For this reason, this theory can only be treated as a low-energy truncation of a fundamental theory.

\begin{table}
\begin{center}
\begin{tabular}{|c|c c|c c|c c|c c|}
\hline
Theory of gravity &~~$\alpha_{\bar{\nu}}$ ~~ &~~$\beta_{\bar{\nu}}$ ~~ &~~$\alpha_{\bar{\mu}}$ ~~ &~~$\beta_{\bar{\mu}}$ ~~ &~~ $\alpha_{\nu}$ ~~&~~ $\beta_{\nu}$ ~~&~~ $\alpha_{\mu}$  ~~&~~ $\beta_{\mu}$  \\
\hline
Chern-Simons gravity & zero  & --- & zero & --- & nonzero & 1 & zero & ---  \\
\hline
Ghost-free scalar-tensor gravities & zero  & --- & zero & --- & nonzero & 1 & nonzero & 1  \\
\hline
Symmetric teleparallel equivalence of GR theory & zero  & --- & zero & --- & nonzero & 1 & nonzero & 1  \\
\hline
Ho\v{r}ava-Lifshitz gravity & zero  & --- & nonzero & 0 or 2 or 4 & zero & --- & nonzero & 1 or 3 \\
\hline
Massive gravity & zero  & --- & nonzero & -2 & zero & --- & zero & --- \\
\hline
Double special relativity theory& zero  & --- & nonzero & -2 or 1 & zero & --- & zero & --- \\
\hline
Extra-dimensional theories& zero  & --- & nonzero & -2 or 2 & zero & --- & zero & --- \\
\hline
Noncommutative gravity& zero  & --- & nonzero & -2 or 0 or 2 & zero & --- & zero & --- \\
\hline
Nonlocal gravity& nonzero  & 0 & zero & --- & zero & --- & zero & --- \\
\hline
Time-dependent Planck mass gravity& nonzero  & 0 & zero & --- & zero & --- & zero & --- \\
\hline
$f(T)$ gravity& nonzero  & 0 & zero & --- & zero & --- & zero & --- \\
\hline
\end{tabular}
\end{center}
\caption{The values of parameter set ($\alpha_{\bar{\nu}}$,$\beta_{\bar{\nu}}$,$\alpha_{\bar{\mu}}$,$\beta_{\bar{\mu}}$,$\alpha_{\nu}$,$\beta_{\nu}$,$\alpha_{\mu}$,$\beta_{\mu}$) defined in Eq.(\ref{coes}) in various modified gravities}
\label{tab1}
\end{table}

\subsection{Ghost-free scalar-tensor gravities}

To avoid the Ostrogradsky ghost in the framework of scalar-tensor gravity with parity violation, in \cite{ghost1} the authors proposed an extension of CS modified gravity by considering the terms which involve the derivatives of scalar field. If considering only the scalar field and its first derivatives in PV terms, the Lagrangian density $\mathcal{L}_{\rm PV}$ is given by
\bqn\label{lv1}
\mathcal{L}_{\rm PV} &=& \sum_{\rm A=1}^4  a_{\rm A}(\phi, \phi^\mu \phi_\mu) L_{\rm A},
\eqn
where $\phi^\mu \equiv \nabla^\mu \phi$, and $a_{\rm A}$ are a priori arbitrary functions of $\phi$ and $\phi^\mu \phi_\mu$. The terms $L_{\rm A}$ are \cite{ghost1,gao}
\[
L_1 = \varepsilon^{\mu\nu\alpha \beta} R_{\alpha \beta \rho \sigma} R_{\mu \nu\; \lambda}^{\; \; \;\rho} \phi^\sigma \phi^\lambda,~
L_2 =  \varepsilon^{\mu\nu\alpha \beta} R_{\alpha \beta \rho \sigma} R_{\mu \lambda }^{\; \; \;\rho \sigma} \phi_\nu \phi^\lambda,~
L_3 = \varepsilon^{\mu\nu\alpha \beta} R_{\alpha \beta \rho \sigma} R^{\sigma}_{\;\; \nu} \phi^\rho \phi_\mu,~
L_4 =  \varepsilon^{\mu\nu\rho\sigma} R_{\rho\sigma \alpha\beta} R^{\alpha \beta}_{\;\;\;\; \mu\nu} \phi^\lambda \phi_\lambda.
\]
In order to avoid the Ostrogradsky modes in the unitary gauge (where the scalar field depends on time only), it is required that $4a_1+2 a_2+a_3 +8 a_4=0$. With this condition, the Lagrangian in Eq.(\ref{lv1}) does not have any higher order time derivative of the metric, but only higher order space derivatives. In this theory, the tensor quadratic action is that in Eq.(\ref{gw-action}) with the coefficients $c_1$, $c_2$ as follows \cite{qiao},
\bqn
{ c_1}/{\red{M_{\rm PV}}} &=& -4 \dot{a_1}\dot{\phi}^2 -8 a_1\dot{\phi}\ddot{\phi} + 8a_1 H \dot{\phi}^2- 2\dot{a_2}\dot{\phi}^2 - 4a_2\dot{\phi}\ddot{\phi}+\dot{a_3}\dot{\phi}^2 +2a_3\dot{\phi}\ddot{\phi} -4a_3 H \dot{\phi}^2-4\dot{a_4}\dot{\phi}^2 -8a_4\dot{\phi}\ddot{\phi}\\
{c_2}/{\red{M_{\rm PV}}} &=&   - 2\dot{a_2}\dot{\phi}^2 -4a_2\dot{\phi}\ddot{\phi} -\dot{a_3}\dot{\phi}^2 -2a_3\dot{\phi}\ddot{\phi} -4\dot{a_4}\dot{\phi}^2 -8a_4\dot{\phi}\ddot{\phi},\label{c2-2}
\eqn
where $H=\dot{a}/a$ is the Hubble parameter. Therefore, the propagation equation of GW becomes that in (\ref{A8}) with \cite{qiao}
\bqn
{\mathcal{H} \bar \nu = 0,~~}
\bar{\mu}=0,~~
{\mathcal{H}\nu_{\rm A} = - \big[ \rho_A c_1 (k/a\red{M_{\rm PV}}) \big] ', ~~ }
\mu_{\rm A}=\rho_{\rm A}(k/a\red{M_{\rm PV}})(c_1-c_2).
\eqn

One can also consider the terms which contains second derivative of the scalar field. Focusing on only the terms those are linear in Riemann tensor and linear/quadratically in the second derivative of $\phi$, the most general Lagrangian $\mathcal{L}_{\rm PV}$ with parity violation is given by \cite{ghost1,gao}
\bqn
\mathcal{L}_{\rm PV} = \sum_{\rm A=1}^7 b_{\rm A} (\phi,\phi^\lambda \phi_\lambda) M_{\rm A},
\eqn
\[
M_1= \varepsilon^{\mu\nu \alpha \beta} R_{\alpha \beta \rho\sigma} \phi^\rho \phi_\mu \phi^\sigma_\nu,~~
M_2= \varepsilon^{\mu\nu \alpha \beta} R_{\alpha \beta \rho\sigma} \phi^\rho_\mu \phi^\sigma_\nu, ~~
M_3= \varepsilon^{\mu\nu \alpha \beta} R_{\alpha \beta \rho\sigma} \phi^\sigma \phi^\rho_\mu \phi^\lambda_\nu \phi_\lambda,
\]
\[
M_4 = \varepsilon^{\mu\nu \alpha \beta} R_{\alpha \beta \rho\sigma} \phi_\nu \phi_\mu^\rho \phi^\sigma_\lambda \phi^\lambda,~~
M_5 = \varepsilon^{\mu\nu \alpha \beta} R_{\alpha \rho\sigma \lambda } \phi^\rho \phi_\beta \phi^\sigma_\mu \phi^\lambda_\nu, ~~
M_6 = \varepsilon^{\mu\nu \alpha \beta} R_{\beta \gamma} \phi_\alpha \phi^\gamma_\mu \phi^\lambda_\nu \phi^\lambda, ~~
M_7 = (\nabla^2 \phi) L_1,
\]
with $\phi^{\sigma}_\nu \equiv \nabla^\sigma \nabla_\nu \phi$. Similarly, in order to avoid the Ostrogradsky modes in the unitary gauge, the following conditions should be imposed: $b_7=0$, $b_6=2(b_4+b_5)$ and $b_2=-A_*^2(b_3-b_4)/2$, where $A_*\equiv \dot{\phi}(t)/N$ and $N$ is the lapse function. After tedious calculations, we find that in this theory, the tensor quadratic action is also given by Eq.(\ref{gw-action}), where $c_2=0$ and $c_1$ is \cite{qiao}
\bqn
{ c_1}/{\red{M_{\rm PV}}} &=& -2 b_1\dot{\phi}^3+4b_2\left(2 H\dot{\phi}^2-\dot{\phi}\ddot{\phi}\right)+2b_3\left(\dot{\phi}^3\ddot{\phi}- H \dot{\phi}^4\right)+2b_4\left(\dot{\phi}^3\ddot{\phi}- H \dot{\phi}^4\right)-2b_5 H \dot{\phi}^4+2b_7\dot{\phi}^3\ddot{\phi}.
\eqn
The propagation equation of GW is derived by varying the quadratic action with respect to $h_{ij}$, which is that in Eq.(\ref{A8}) with \cite{qiao}
\bqn
{\mathcal{H} \bar \nu = 0,~~}
\bar{\mu}=0,~~
{\mathcal{H}\nu_{\rm A} = - \big[ \rho_A c_1 (k/a\red{M_{\rm PV}}) \big] ', ~~ }
\mu_{\rm A}=\rho_{\rm A}(k/a\red{M_{\rm PV}})c_1.
\eqn


\subsection{Symmetric teleparallel equivalence of GR theory}

{{Parity-violating extension of the symmetric teleparallel equivalent of GR theory}} is a non-Riemannian formulation of gravity, which allows for a wider variety of consistent extensions than the metric formulation of gravity theory, and has been studied explicitly in \cite{tele}. As known, the Einstein-Hilbert action hides second derivatives and its generalizations are severely restricted. However, combining the Riemannian formulation with the metric teleparallel \cite{tele18}, the symmetric teleparallel \cite{tele19}
equivalents of GR provides a more flexible framework for generalizations, since their action principles can feature only first derivatives. For this reason, the analysis in the subsection is beyond the framework in Sec. \ref{sec2}.

The non-metricity tensor and its contractions are defined by $Q_{\alpha\mu\nu}\equiv \tilde{\nabla}_{\alpha} g_{\mu\nu}$, where the covariant derivative $\tilde{\nabla}$ is with respect to a generic connection $\Gamma$, which is considered to be devoid of both curvature and torsion. This allows us to choose the coincident gauge \cite{tele20} wherein we can simply write partial derivatives in place of the covariant operators above. The action of the non-metricity equivalent of GR $S_{\rm QGR}$ is given in Eq.(2.15) in \cite{tele}, which is related to the Einstein-Hilbert action by a total derivative. In this article, we investigate the PV gravity that is quadratic in non-metricity in the symmetric teleparallel geometry, and consider only the models that feature (at most) second time-derivatives. Let us concentrate on the first class of Lagrangians, which are quadratic
in non-metricity, coupled quadratically to a scalar field and second-order in derivatives. The unique and non-vanishing Lagrangian for GW is given by \cite{tele}
\beqa
\mathcal{L}_{\rm PV}=\sqrt{-g} \tilde{\alpha} \epsilon^{\mu\nu\theta\delta}\phi_{\theta}\phi^{\gamma}Q_{\mu\nu\alpha}Q_{\gamma\delta}^{~~\alpha},
\eeqa
where $\tilde{\alpha}$ is an arbitrary function of field $\phi$ and its kinetic term. For the propagation of GW, we need only consider the tensor quadratic action, which is
\beqa
\mathcal{L}^{(2)}=\mathcal{L}^{(2)}_{\rm QGR}+\mathcal{L}^{(2)}_{\rm PV},
\eeqa
where
\beqa
\mathcal{L}^{(2)}_{\rm QGR}=\frac{1}{4}\left[\dot{h}_{ij}^2-a^{-2}(\partial_{k} h_{ij})^2\right],~~
\mathcal{L}^{(2)}_{\rm PV}=\frac{H}{a}\tilde{\alpha}\epsilon^{ijk}h_k^{~l}\partial_i h_{jl}.
\eeqa
From this action, the propagation equation of GW is derived, that in Eq.(\ref{A8}) with \cite{tele}
\bqn
{\mathcal{H} \bar \nu=0, ~~}
\bar{\mu}=0,~~
{ \mathcal{H} \nu_{\rm A}=0, ~~}
\mu_{\rm A}=4\tilde{\alpha}\rho_{\rm A}\mathcal{H}/k.
\eqn

In a similar fashion, we can discuss the PV Lagrangians in the symmetric teleparallel geometry with three-order derivatives. Let us focus on the action, which is quadratic in non-metricity, and includes (at most) the second order time derivatives. The most general Lagrangian density is given by \cite{tele}
\beqa
\mathcal{L}^{(2)}=\mathcal{L}^{(2)}_{\rm QGR}+
\frac{\beta_1(t)}{a^3\red{M_{\rm PV}}}\mathcal{L}^{(2)}_{\rm PV1}+
\frac{\beta_2(t)}{a\red{M_{\rm PV}}}\mathcal{L}^{(2)}_{\rm PV2}+
\frac{\beta_3(t)}{a\red{M_{\rm PV}}}\mathcal{L}^{(2)}_{\rm PV3},
\eeqa
where $\beta_i(t)$ is a dimensionless function of time, $\red{M_{\rm PV}}$ is an energy scale and
\beqa
\mathcal{L}^{(2)}_{\rm PV1}\equiv \epsilon^{ijk}\partial^2 h_j^{~l}\partial_i h_{kl},~~
\mathcal{L}^{(2)}_{\rm PV2}\equiv 2{H}\epsilon^{ijk}\dot{h}_j^{~l}\partial_i h_{kl},~~
\mathcal{L}^{(2)}_{\rm PV3}\equiv \epsilon^{ijk}\dot{h}_j^{~l}\partial_i \dot{h}_{kl}.
\eeqa
Note that, in the symmetric teleparallel equivalence of GR theory, the extra term $\mathcal{L}^{(2)}_{\rm PV2}$ can exist. Similarly, the propagation equation of GW is derived from the action, that in Eq.(\ref{A8}) with \cite{tele}
\beqa
{ \mathcal{H} \bar{\nu}=0,~~}
\bar{\mu}=0,~~
{\mathcal{H} \nu_{\rm A}=  \big[ 4 \rho_A(k/a\red{M_{\rm PV}}) \beta_3 \big] ',~~}
\mu_{\rm A}={ 4 \rho_A(k/a\red{M_{\rm PV}})(\beta_1-\beta_3+ {(a \mathcal{H}\beta_2)'}/{a k^2} )}.
\eeqa
Note that, to avoid fine-tuning, $\beta_1$, $\beta_2$ and $\beta_3$ are expected to be of the same order. Thus, the last term in $\mu_{\rm A}$ become ignorable.

\subsection{Ho\v{r}ava-Lifshitz gravity}



The Ho\v{r}ava-Lifshitz (HL) gravity is based on the perspective that the Lorentz symmetry appears only as an emergent symmetry at low energies, but can be fundamentally absent at high energies \cite{Horava:2009uw,Wang:2017brl}. This opens a completely new window to build a theory of quantum gravity without the Lorentz symmetry in the UV, using the high-dimensional spatial derivative operators, while still keeping the time derivative operators to the second-order, whereby the unitarity of the theory is reserved. Besides the original version of the theory by Ho\v{r}ava \cite{Horava:2009uw}, there are several modifications, which are absent several in-consistent problems that appear in the original version. In this paper, we are going to focus on an extension of the HL gravity by abandoning the projectability condition but imposing an extra local U(1) symmetry that was proposed \cite{Zhu:2011xe,zhu2012}, in which the gravitational sector has the same degree of freedom as that in GR, i.e., only spin-2 massless gravitons exist.

By abandoning the Lorentz symmetry, the HL theory also provide a natural way to incorporate the parity violation terms into the theory. For our current purpose, we consider the third- and/or fifth-order spatial derivative operators to the potential term ${\cal{L}}$ of the total action in \cite{Zhu:2011xe,zhu2012,wang_polarizing_2012,zhu_effects_2013},
\beqa \label{parity action}
 {\cal{L}_{\rm PV}} &=& \frac{1}{\red{M_{\rm PV}}^3} \left(\alpha_0 K_{ij} R_{ij} +\alpha_2 \varepsilon^{ijk} R_{il} \Delta_j R^l_k \right) + \frac{\alpha_1 \omega_3(\Gamma)}{\red{M_{\rm PV}}}+``\dots".
\eeqa
Here $\red{M_{\rm PV}}$ is the energy scale above which the high-order derivative operators become important. The coupling constant $\alpha_0,\;\alpha_1,\;\alpha_2$ are dimensionless and arbitrary, and $\omega_3(\Gamma)$  the 3-dimensional gravitational CS term.  ``..." denotes the rest of the fifth-order operators given in Eq.(2.6) of \cite{zhu2012}. Since they have no contributions to tensor perturbations, in this paper we shall not write them out explicitly. 


The general formulas of the linearized tensor perturbations were given in \cite{zhu_effects_2013}, so in the rest of this section we give a very brief summary of  the main results obtained there.  Consider a flat FRW universe and assuming that matter fields have no contributions to tensor perturbations, the quadratic part of the total action can be cast in the form,
\beqa
&&S^{(2)}_{\text{ g}}=\zeta^2 \int d\eta d^3 x \Bigg\{\frac{a^2}{4} (h_{ij}')^2-\frac{1}{4} a^2 (\partial_k h_{ij})^2-\frac{\hat{\gamma_3}}{4 \red{M_{\rm PV}}^2}(\partial^2h_{ij})^2-\frac{\hat{\gamma}_5}{4\red{M_{\rm PV}}^4 a^2}(\partial^2 \partial_k h_{ij})^2\nonumber\\
&&\;\;\;\;\;\;\;-\frac{\alpha_1 a \epsilon^{ijk}}{2 \red{M_{\rm PV}}} (\partial_l h_{i}^m \partial_m \partial_j h_k^l-\partial_l h_{im} \partial^l \partial_j h^m_k)-\frac{\alpha_2 \epsilon^{ijk}}{4\red{M_{\rm PV}}^3 a}\partial^2 h_{il} (\partial^2 h^l_k)_{,j} - \frac{3 \alpha_0 {\cal{H}}}{8 \red{M_{\rm PV}} a }(\partial_k h_{ij})^2\Bigg\},
\eeqa
where $\hat{\gamma}_3  \equiv  ({2\red{M_{\rm PV}}}/{M_{\rm Pl}})^2  {\gamma}_3$ and $\hat{\gamma}_5 \equiv({2\red{M_{\rm PV}}}/{M_{\rm Pl}})^4 {\gamma}_5$, and $\gamma_3$ and $\gamma_5$ are the dimensionless coupling constants of the theory. To avoid fine-tuning, ${{\alpha}}_{n}$ and  $\hat{\gamma}_{n}$  are expected to be  of  the same order.
Then, the field equations for $h_{ij}$ read,
 \beqa
 \label{DFE}
 h''_{ij}&+&2\mathcal{H}h'_{ij}- \alpha^2 \partial^2
 h_{ij}+ \frac{\hat\gamma_3}{a^2\red{M_{\rm PV}}^2}\partial^4
 h_{ij}-\frac{\hat\gamma_5}{a^4\red{M_{\rm PV}}^4}\partial^6 h_{ij} +
e_{i}^{\;\; lk}\left(\frac{2\alpha_1}{a\red{M_{\rm PV}}} + \frac{\alpha_2}{a^3\red{M_{\rm PV}}^3}\partial^2 \right) \left(\partial^2h_{jk}\right)_{,l} = 0,
 \label{eq7}
 \eeqa
where $\alpha^2 \equiv 1+ {3\alpha_0{\cal{H}}}/{(2\red{M_{\rm PV}} a)}$. In the late universe, $a\sim 1$, and $\mathcal{H}\ll \red{M_{\rm PV}}$, so we find $\alpha^2\rightarrow 1$. To study the evolution of $h_{ij}$, we expand it over spatial Fourier harmonics. For each circular polarization mode, the equation of motion of GW is given by
 \beqa
 h_{\rm A}'' +2 \mathcal{H} h_{\rm A}'+\omega_{\rm A}^2 h_{\rm A}=0,
 \eeqa
 with
\beqa
 \omega^2_{\rm A}(k, \eta) &\equiv& \alpha^2 k^2 \left[1+ \delta_1 \rho_{\rm A} \left({\alpha k}/{\red{M_{\rm PV}} a}\right)+\delta_2 \left({\alpha k}/{\red{M_{\rm PV}} a}\right)^2 - \delta_3 \rho_{\rm A} \left({\alpha k}/{\red{M_{\rm PV}} a}\right)^3 + \delta_4 \left({\alpha k}/{\red{M_{\rm PV}} a}\right)^4 \right],
 \eeqa
where $\delta_1\equiv {2 \alpha_1 }/{\alpha^3 }$,
$\delta_2\equiv {\hat{\gamma_3} }/{\alpha^4 }$,
 $\delta_3\equiv {\alpha_2 }/{\alpha^5 }$,
 $\delta_4\equiv {\hat{\gamma_5} }/{\alpha^6 }$. In comparison with the formula in Eq.(\ref{A8}), we find the coefficients in HL gravity as follows
 \beqa
{ \mathcal{H} \bar{\nu}=0,~~}
 \bar{\mu}=\delta_2 \left(k/a\red{M_{\rm PV}}\right)^2+\delta_4 \left(k/a\red{M_{\rm PV}}\right)^4+{3\alpha_0{\cal{H}}}/{(2\red{M_{\rm PV}} a)},~~
 {\mathcal{H} \nu_{\rm A}=0,~~}
 \mu_{\rm A}=\delta_1 \rho_{\rm A} \left(k/a\red{M_{\rm PV}}\right)- \delta_3 \rho_{\rm A} \left(k/a\red{M_{\rm PV}}\right)^3,\nb\\
 \eeqa
where we have considered the relation $\alpha^2\rightarrow1$. In the expression of $\bar{\mu}$, the second term is always negligible, and the relative magnitude of the first and third terms depends on the values of $k$ and $\red{M_{\rm PV}}$. In the theory, which includes both the third- and fifth-order operators \cite{wang_polarizing_2012,zhu_effects_2013}, the first term in $\mu_{\rm A}$ is dominant. While for the theory, which includes only the fifth-order operator \cite{soda}, only the second term in $\mu_{\rm A}$ exists.

\subsection{Various theories of gravity without parity violation}

It is important to mention that, although in this paper we focus on GWs in the parity-violating gravities, the equation of motion of GW in Eq.(\ref{A8}) can also include various cases in the parity-conserving gravities. In the most general scalar-tensor gravities, i.e. the Horndeski, beyond Horndeski gravities and degenerate higher-order scalar-tensor theories, the propagation equations of GW correspond to Eq.(\ref{A8}) with nonzero $\bar\nu$ and $\bar\mu$ \cite{gao,horn,lorentz2,dhost}.

Another case is the theories of gravity with modified dispersion relations, the propagation equation of GW is Eq.(\ref{A8}), where the parameters $\bar\nu$, $\nu_{\rm A}$, $\mu_{\rm A}$ are zero, but $\bar\mu$ is nonzero. A few examples of such modified dispersion relations include the following \cite{will}:

$\bullet$~{\emph{Massive gravity}} has the dispersion relation $E^2=p^2+m_g^2$ with $m_g$ the graviton's rest mass \cite{will,massive}, which follows the equation of motion of GW in Eq.(\ref{A8}) with nonzero $\bar\mu=(m_g/k)^2$.

$\bullet$~{\emph{Double special relativity theory}} has the relation $E^2=p^2+m_g^2+\eta_{\rm dsrt}E^3+\cdot\cdot\cdot$ with $\eta_{\rm dsrt}$ a parameter of the order of Planck length \cite{34-37}, which follows the Eq.(\ref{A8}) with nonzero $\bar\mu=(m_g/k)^2+\eta_{\rm dsrt} k +\cdot\cdot\cdot$.

$\bullet$~{\emph{Extra-dimensional theories}} have the relation $E^2=p^2+m_g^2-\alpha_{\rm edt}E^4$ with $\alpha_{\rm edt}$ a constant related to the square of the Planck length \cite{38}, which follows the Eq.(\ref{A8}) with nonzero $\bar\mu=(m_g/k)^2-\alpha_{\rm edt}k^2$.

$\bullet$~{\emph{\red{Noncommutative gravity}}}  \footnote{In noncommutative gravity, the spacetime coordinates are promoted to operators, which satisfy the nontrivial commutation relations $[\hat x^\mu, \hat x_\nu] = i \Theta^{\mu\nu}$ with $\Theta^{\mu\nu}$ being a real constant antisymmetric tensor.} has the relation $E^2 g_1^2(E)=m_g^2+p^2$ with $g_1=(1-\sqrt{\alpha_{\rm ncg}\pi}/2)\exp(-\alpha_{\rm ncg}E^2/E_p^2)$ with $\alpha_{\rm ncg}$ a constant \cite{43-45}, which follows the Eq.(\ref{A8}) with nonzero $\bar\mu=(1+\sqrt{\alpha_{\rm ncg}\pi}/2)(1+(m_g/k)^2+\alpha_{\rm ncg} (k/E_p)^2+(m_g/E_p)^2)$.

In some theories of parity-conserving gravity, the propagation of GW is described by Eq.(\ref{A8}) with nonzero $\bar{\nu}$, but the other parameters $\bar\mu$, $\mu_{\rm A}$ and $\nu_{\rm A}$ are all zero. One example is the nonlocal gravity \cite{nonlocal}, in which the action of gravity is given by
\[
S=\frac{1}{16\pi G} \int d^4 x \sqrt{-g} \left[R-\frac{1}{6}m^2\frac{1}{\Box}R\right],
\]
where $m$ is a mass parameter that replaces the cosmological constant. The motion of equation of GW in this theory is given by Eq.(\ref{A8}) with
$\mathcal{H}\bar\nu=-[3\bar{V}/(2-6\gamma\bar{V})]'$,
where $\gamma=m^2/9H_0^2$ and $\bar{V}$ is the background evolution of an auxiliary field in the theory \cite{nonlocal,57}. In these theories, due to the modified friction term in the perturbation equation of cosmological GWs, the damping of GW amplitude with the expansion of the universe is different from that in GR. As a result, the effective luminosity distance of GW is different from that of electromagnetic wave \cite{nonlocal}. Similarly, in the theory of gravity with the time-dependent effective Planck mass, the motion of equation of GW is also given by Eq.(\ref{A8}) with $\mathcal{H}\bar\nu=[\ln M_*^2]'$ with $M_*$ the time-dependence Planck mass \cite{effective-planck-mass}. Recently, the propagation of GW in the $f(T)$ gravitational theory is also studied explicitly in \cite{cai}, and find the equation of motion have the same form with $\mathcal{H}\bar\nu=[\ln f_{T}]'$, where $T$ is the torsion scalar defined in the theory, $f(T)$ is the arbitrary function of $T$, and $f_{T}\equiv df(T)/dT$.


\section{Amplitude and velocity birefringences \label{sec4}}
\renewcommand{\theequation}{4.\arabic{equation}} \setcounter{equation}{0}

In this section, in the general theories of PV gravity, we study the velocity and amplitude birefringence effects during the propagation of GWs. As discussed in Sec. \ref{sec2}, the equation of motion of GWs is Eq.(\ref{A8}), and the parameters are give by Eq.(\ref{coes}). For each circular polarization mode $h_{\rm A}$, the velocity and amplitude birefringence effects induce the the phase and amplitude corrections to the waveform of GWs. Similar to the previous work \cite{qiao}, to investigate these two effect separately, we decompose $h_{\rm A}$ as
\bqn
h_{\rm A} = {\bar h}_{\rm A} e^{-i \theta(\tau)}, ~~~~
{\bar h}_{\rm A} = \mathcal{A}_{\rm A} e^{- i \Phi(\tau)},\lb{decom1}
\eqn
where ${\bar h}_{\rm A}$ satisfies
\bqn\lb{hAbar}
{\bar h}_{\rm A} '' + 2 \mathcal{H} {\bar h}'_{\rm A} + (1+\bar{\mu}+\mu_{\rm A}) {k^2} {\bar h}_{\rm A}=0.
\eqn
Here $\mathcal{A}_{\rm A}$ and $\Phi(\tau)$ are the amplitude and phase of ${\bar h}_{\rm A}$ respectively. With this decomposition, $\theta (\tau)$ encodes the correction arising from $\bar{\nu}$ and $\nu_A$, while the corrections due to $\bar{\mu}$ and $\mu_A$ are included in ${\bar h}_{\rm A}$. \red{In the ghost-free PV gravities, since $\bar \nu=0 = \bar \mu$, the phase $\theta(\tau)$ only contains corrections from the parity violation, which is directly encoded by $\nu_A$, while the amplitude correction from $\mu_A$ is included in $\bar h_A$, as one can see from Eq. (4.5) in \cite{qiao}. In the generic case, the phase $\theta(\tau)$ and the amplitude $\bar h_A$ can receive corrections arising from the modifications of gravity which are not relevant to the parity violation. As we mentioned, their effects are encoded by the parameters $\bar \mu$ and $\bar \nu$. }

\subsection{Phase modifications}
We first concentrate on the corrections arising from the parameters $\mu$ and $\mu_A$, which modify the dispersion relations of GWs. To proceed, we define ${\bar u}_{\rm A}(\tau) = \frac{1}{2} a(\tau) M_{\rm Pl}{\bar{h}}_{\rm A} (\tau) $ and then Eq. (\ref{hAbar}) can be written as
\bqn
\frac{d^2 {\bar u}_{\rm A}}{d\tau^2} + \left(\omega_{\rm A}^2 - \frac{a''}{a}\right) {\bar u}_{\rm A}=0,
\eqn
where
\bqn
\omega_{\rm A}^2 = k^2(1+\bar{\mu}+\mu_{\rm A})=k^2\left(1+\alpha_{\bar{\mu}}(\tau)(k/a\red{M_{\rm PV}})^{\beta_{\bar{\mu}}}+\rho_{\rm A}\alpha_{\mu}(\tau)(k/a\red{M_{\rm PV}})^{\beta_{\mu}}\right),
\eqn
is the modified dispersion relation. With this relation, the speed of the graviton reads
${v_{\rm A}^2}/{c^2} \simeq 1-\bar{\mu}-\mu_{\rm A}$,
which leads to
\bqn
{v_{\rm A}}/{c} \simeq 1- (1/2)\alpha_{\bar{\mu}}(\tau)(k/a\red{M_{\rm PV}})^{\beta_{\bar{\mu}}}-({1}/{2}) \rho_{\rm A}  \alpha_\mu(\tau) \left( {k}/{a\red{M_{\rm PV}}}\right)^{\beta_\mu}.
\eqn
We find that, the parameter $\bar{\mu}$ influences both circular modes in a same way, which has been tightly constrained by comparing the arrival times of GW170817 and GRB170817a \cite{gw170817-speed}. However, for the parameter $\mu_{\rm A}$, since $\rho_{\rm A}$ have the opposite signs for left-hand and right-hand polarization modes, its effects on two modes are opposite, which induces the velocity birefringence effect of GWs. For this reason, in order to test the parity-violating effect by the velocity of GWs, we should measure the velocity difference of two circular polarization modes, instead of just measuring the speed of an individual mode \cite{zhao2019}.

Consider gravitons with same $\rho_{\rm A}$ emitted at two different times $t_e$ and $t_e'$, with wave numbers $k$ and $k'$, and received at corresponding arrival times ($r_e$ is the same for both). Assuming $\Delta t_e\equiv t_e-t_e'\ll a/\dot{a}$, the difference of arrival times of these two waves are given by \cite{will,qiao,zhao2019}
\bqn\label{deltat0}
\Delta t_0 = (1+z)\Delta t_e+ \frac{\rho_{\rm A}}{2}\left( (\frac{k}{ \red{M_{\rm PV}}})^{\beta_\mu}-(\frac{k'}{ \red{M_{\rm PV}}})^{\beta_\mu}\right) \int_{t_e}^{t_0} \frac{\alpha_{\mu}}{a^{\beta_\mu+1}}dt
+ \frac{1}{2}\left( (\frac{k}{ \red{M_{\rm PV}}})^{\beta_{\bar\mu}}-(\frac{k'}{ \red{M_{\rm PV}}})^{\beta_{\bar\mu}}\right) \int_{t_e}^{t_0} \frac{\alpha_{\bar\mu}}{a^{\beta_{\bar\mu}+1}}dt,
\eqn
where $z\equiv 1/a(t_e)-1$ is the cosmological redshift. Let us focus on the GW signal generated by non-spinning, quasi-circular inspiral in the post-Newtonian approximation. Relative to the GW in GR, the terms $\bar\mu$ and $\mu_{\rm A}$ modify the phase of GW $\Phi(\tau)$. The Fourier transform of $\bar{h}_{\rm A}$ can be obtained analytically in the stationary phase approximation, where we assume that the phase is changing much more rapidly than the amplitude, which is given by \cite{spa}
\bqn
\tilde{\bar{h}}_{\rm A}(f)=\frac{{\mathcal{A}}_{\rm A}(f)}{\sqrt{df/dt}}e^{i\Psi(f)},
\eqn
where $f$ is the GW frequency at the detector, and $\Psi$ is the phase of GWs. In \cite{will}, it was proved that, the difference of arrival times in Eq.(\ref{deltat0}) induces the modification of GWs phases $\Psi$ as follows,
\bqn
\Psi_{\rm A}(f) = \Psi_{\rm A}^{\rm GR} (f) + \rho_{\rm A}\delta \Psi_{1}(f)+\delta \Psi_{2}(f).
\eqn
When $\beta_{\mu}\neq -1$,
\bqn \label{deltapsi1-1}
\delta \Psi_{1}(f) = \frac{(2/\red{M_{\rm PV}})^{\beta_\mu}}{\beta_{\mu}+1}\frac{u^{\beta_{\mu}+1}}{{\mathcal{M}}^{{\beta_\mu}+1}}\int_{t_e}^{t_0} \frac{\alpha_{\mu}}{a^{\beta_\mu+1}}dt,
\eqn
and when $\beta_{\mu}= -1$,
\bqn\label{deltapsi1-2}
\delta \Psi_{1}(f)=\frac{\red{M_{\rm PV}}}{2} \ln u \int_{t_e}^{t_0} {\alpha_{\mu}}dt.
\eqn
Similarly, since $\beta_{\bar\mu}$ is an even number, we have
\bqn \label{deltapsi2-1}
\delta \Psi_{2}(f) =
\frac{(2/\red{M_{\rm PV}})^{\beta_{\bar\mu}}}{\beta_{\bar\mu}+1}\frac{u^{\beta_{\bar\mu}+1}}{{\mathcal{M}}^{{\beta_{\bar\mu}}+1}}\int_{t_e}^{t_0} \frac{\alpha_{\bar\mu}}{a^{\beta_{\bar\mu}+1}}dt.
\eqn
Here, we have defined $u=\pi \mathcal{M} f$. The quantity $\mathcal{M} = (1+z) \mathcal{M}_{\rm c}$ is the measured chirp mass, and $\mathcal{M}_{\rm c}\equiv (m_1 m_2)^{3/5}/(m_1+m_2)^{1/5}$ is the chirp mass of the binary system with component masses $m_1$ and $m_2$. Note that, the phase modification in Eqs.(\ref{deltapsi1-1})-(\ref{deltapsi2-1}) is a simple propagation effect, which is independent of the  generation effect of GWs. Although this formula is obtained in the stationary phase approximation \cite{will}, we expect this result is also applicable for the GWs in the more general cases in the presence of spin, precession, eccentricity of compact binaries, and/or for the GWs produced during the merger and ring-down of compact binaries. This extension has been adopted by LIGO and Virgo collaborations in \cite{gw150914,gw170817-speed}.

\subsection{Amplitude modifications}

Now, let us turn to study the effects caused by $\bar{\nu}$ and $\nu_A$. Plugging the second equation of the decomposition (\ref{decom1}) into (\ref{hAbar}), one finds the equation for $\Phi(t)$,
\bqn\lb{eom_Phi}
i \Phi'' + \Phi'^2 + 2 i \mathcal{H} \Phi' - (1+\bar{\mu}+\mu_{\rm A})k^2=0.
\eqn
Similarly, plugging the first equation of the decomposition (\ref{decom1}) into (\ref{A8}), one obtains
\bqn \lb{eom2}
&&i (\theta''+\Phi'') + (\Phi'+\theta')^2+i (2+\bar{\nu}+\nu_{\rm A})\mathcal{H} (\theta'+\Phi') - (1+\bar{\mu}+\mu_{\rm A})k^2=0.
\eqn
Using the equation of motion (\ref{eom_Phi}) for $\Phi$, the above equation reduces to
\bqn
i \theta''+ 2 \theta' \Phi' + \theta'^2 + i (2+\bar{\nu}+\nu_{\rm A})\mathcal{H} \theta'+ i (\bar{\nu}+\nu_{\rm A}) \mathcal{H} \Phi'=0.
\eqn
The phase $\Phi$ is expected to be close to that in GR $\Phi_{\rm GR}$, and $\Phi_{\rm GR}'  \sim k$, where the wave number relates to the GW frequency by $k=2\pi f/a_0$. {\color{black} Since the amplitude modification function $\theta$ is caused by the expansion of the universe, we have $\theta'\sim \mathcal{H}$ and $\theta''\sim \mathcal{H}^2$. Note that, the conformal Hubble parameter $\mathcal{H}\sim 10^{-18}$Hz, which is much smaller than the GW frequency ($\sim 10^2$Hz) of compact binaries. Considering that $\theta'' \ll \Phi'\theta'  \sim k \theta'$, $k \gg \mathcal{H}$,
and keeping only the leading-order terms}, the above equation can be simplified into the form $2  \theta' + i \mathcal{H} (\bar{\nu}+\nu_{\rm A})  =0$,
which has the solution
\bqn
 \theta = -\frac{i}{2} \int_{\tau_e}^{\tau_0} \mathcal{H} (\bar{\nu}+\nu_{\rm A}) d\tau.
 \eqn
We observe that the contributions of $\bar{\nu}$ and $\nu_{\rm A}$ in the phase is purely imaginary. This indicates that these parameters lead to modification of the amplitude of the GWs during the propagation. As a result, relative to the corresponding mode in GR, the amplitude of left-hand circular polarization of GWs will increase (or decrease) during the propagation, while the amplitude for the right-hand mode will decrease (or increase).

More specifically, for the parameters $\bar{\nu}$ and $\nu_{\rm A}$ given in Eq. (\ref{coes}), one can write the waveform of GWs with parity violation effects in the form
\bqn\lb{waveformA}
h_{\rm A} = h_{\rm A}^{\rm GR} (1+\rho_{\rm A}\delta h_1+\delta h_2) e^{ - i \delta \Phi_{\rm A}},
\eqn
where
\bqn\lb{waveform}
{\delta h_1 =-\frac{1}{2} \left[a_\nu \left(\frac{k}{a \red{M_{\rm PV}}}\right)^{\beta_{\nu}} \right]\Bigg|_{a_e}^{a_0} ,~~
\delta h_2 =-\frac{1}{2} \left[a_{\bar \nu} \left(\frac{k}{a \red{M_{\rm PV}}}\right)^{\beta_{\bar \nu}} \right]\Bigg|_{a_e}^{a_0}}
\eqn
Using the notations $u$ and $\mathcal{M}$, one can rewrite $\delta h_1$ and $\delta h_2$ in the forms
{\bqn\label{deltah12}
\delta h_1 = -\frac{1}{2} \left(\frac{2u}{\red{M_{\rm PV}}\mathcal{M}}\right)^{\beta_\nu} \Big[ a_\nu(\tau_0) - a_\nu(\tau_e) (1+z)^{\beta_\nu}\Big], ~~
\delta h_2 = -\frac{1}{2} \left(\frac{2u}{\red{M_{\rm PV}}\mathcal{M}}\right)^{\beta_{\bar\nu}} \Big[ a_{\bar \nu}(\tau_0) - a_{\bar \nu}(\tau_e) (1+z)^{\beta_{\bar \nu}}\Big].
\eqn}

\subsection{GW waveforms of the circular polarization modes}

For each circular polarization mode, we summarize the modification of GW waveforms in this subsection. Similar to the above discussion, we consider the GWs produced during the inspirlling stage of the compact binaries. In this scheme, in the stationary phase approximation, the Fourier transform of $h_{\rm A}$ can be calculated directly, which is given by (similar to the parameterized post-Einsteinian formulae in \cite{ppe})
\bqn\label{hA-f}
\tilde{h}_{\rm A}(f)=\tilde{h}_{\rm A}^{\rm GR}(f) (1+\rho_{\rm A}\delta h_1+\delta h_2) e^{i(\rho_{\rm A}\delta\Psi_1+\delta\Psi_2)},
\eqn
where $\tilde{h}_{\rm A}^{\rm GR}(f)$ is the corresponding waveform in GR, and the explicit formula can be found in the previous works \cite{pn,spa}. Here, we focus on the correction terms on both the amplitude and phase of GWs. The former ones are caused by the parameters $\bar{\nu}$ and $\nu_{\rm A}$ in Eq.(\ref{A8}), while the latter ones are caused by the parameters $\bar{\mu}$ and $\mu_{\rm A}$. Since in the theories of PV gravity, the GW propagation equations of left-hand and right-hand polarization modes are decoupled, the parameters $\nu_{\rm L}$ and $\mu_{\rm L}$ ($\nu_{\rm R}$ and $\mu_{\rm R}$) can only affect the amplitude and phase of the left-hand (right-hand) polarization mode. Due to the opposite signs of $\nu_{\rm L}$ and $\nu_{\rm R}$ (as well as $\mu_{\rm L}$ and $\mu_{\rm R}$), they induce the birefringence effects of GWs. However, the parameters $\bar{\nu}$ and $\bar{\mu}$ influence the amplitudes and phases of two modes in a same manner. Therefore, as expected, they are not relevant to parity violation in gravity, and cannot cause the GW birefringence effects.

\section{Modifications to the GW waveform \label{sec5}}
\renewcommand{\theequation}{5.\arabic{equation}} \setcounter{equation}{0}

In order to make contact with observations, it is convenient to analyze the GWs in the Fourier domain, and the responses of detectors for the GW signals $\tilde{h}(f)$ can be written in terms of waveforms of $\tilde{h}_+$ and $\tilde{h}_\times$ as
\bqn
\tilde{h}(f) = [F_+ \tilde{h}_+(f) + F_\times \tilde{h}_\times(f)] e^{- 2 \pi i f \Delta t},
\eqn
where $F_{+}$ and $F_{\times}$ are the beam pattern functions of GW detectors, depending on the source location and polarization angle \cite{F+Fx}. $\Delta t$ is the arrival time difference between the detector and the geocenter. In GR, the waveform of the two polarizations $\tilde{h}_+(f)$ and $\tilde{h}_\times (f)$ are given respectively by \cite{pn,spa,300}
\bqn
\tilde{h}^{\rm GR}_+ = (1+\chi^2) \mathcal{A}e^{i \Psi}, \;\;
\tilde{h}^{\rm GR}_\times = 2 \chi \mathcal{A} e^{i (\Psi+\pi/2)},
\eqn
where $\mathcal{A}$ and $\Psi$ denote the amplitude and phase of the waveforms $h^{\rm GR}_{+ \; \times}$, and $\chi=\cos\iota$ with $\iota$ being the inclination angle of the binary system. In GR, the explicit forms of $\mathcal{A}$ and $\Psi$ have been calculated in the high-order PN approximation (see for instance \cite{pn} and references therein). Now we would like to derive how the parity violation in the modified gravities can affect both the amplitude and the phase of the above waveforms. The circular polarization modes $\tilde{h}_{R}$ and $\tilde{h}_L$ relate to the modes $\tilde{h}_+$ and $\tilde{h}_{\times}$ via
\bqn
\tilde{h}_+ =  \frac{\tilde{h}_{\rm L} + \tilde{h}_{\rm R}}{\sqrt{2}},~~
\tilde{h}_\times =  \frac{\tilde{h}_{\rm L} - \tilde{h}_{\rm R}}{\sqrt{2} i}.
\eqn
Similar to the previous work \cite{cs4,qiao}, throughout this paper, we ignore the parity-violating generation effect, which is caused by a modified energy loss, inspiral rate and chirping rate of the binaries. Since the generation effect occurs on a radiation-reaction time scale, which is much smaller than the GW time of flight, making its impact on the evolution of the GW phase negligible \cite{cs3}. In addition, the extra polarization modes of GW can also been produced in the modified gravities. For instance, in the scalar-tensor gravities, the breathing mode and/or longitude mode can be generated \cite{zhao2017,extra} by the coalescence of compact binaries. However, the amplitudes of these extra modes are always much smaller than the tensorial modes $h_+$ and $h_{\times}$, or $h_{\rm R}$ and $h_{\rm L}$, if considering the model constraints from the observations in solar systems, or binary pulsars, and in cosmology \cite{zhao2017,zhang2017}. For these reasons, in this paper, we do not consider this effect as well. Thus, the circular polarization modes $h_{\rm A}$ are given in (\ref{hA-f}), and the waveforms for the plus and cross modes become
\bqn
\tilde{h}_+ &=& \tilde{h}_+^{\rm GR}(1+\delta h_2 +i\delta \Psi_2)-\tilde{h}_{\times}^{\rm GR}(i\delta h_1-\delta \Psi_1) \\
\tilde{h}_\times &=& \tilde{h}_{\times}^{\rm GR}(1+\delta h_2 +i\delta \Psi_2)+\tilde{h}_{+}^{\rm GR}(i\delta h_1-\delta \Psi_1).
\eqn
Therefore, the Fourier waveform $\tilde{h}(f)$ becomes
\bqn\label{final-hf}
\tilde{h}(f)= \mathcal{A} \delta \mathcal{A} e^{i (\Psi +\delta \Psi )} ,
\eqn
where
\bqn\label{final-delta}
\delta \mathcal{A}
&=&\sqrt{(1+\chi^2)^2F^2_+ +4\chi^2F^2_\times}  \left[1+\delta h_2+\frac{2\chi(1+\chi^2)(F^2_+ + F^2_\times)}{(1+\chi^2)^2F^2_+ +4\chi^2F^2_\times}\delta h_1 -\frac{(1-\chi^2)^2F_+F_\times}{(1+\chi^2)^2F^2_+ +4\chi^2F^2_\times}\delta\Psi_1\right],\nb\\
\delta \Psi &=&\tan^{-1}\left[\frac{2\chi F_\times}{(1+\chi^2)F_+}\right]+\delta \Psi_2+\frac{(1-\chi^2)^2 F_+ F_\times}{(1+\chi^2)^2 F^2_+ + 4\chi^2F^2_\times}\delta h_1
+\frac{2\chi(1+\chi^2)(F^2_+ + F^2_\times)}{(1+\chi^2)^2 F^2_+ + 4\chi^2F^2_\times}\delta\Psi_1,
\eqn
where $\delta h_1$ and $\delta h_2$ are given by Eq.(\ref{deltah12}), while $\delta \Psi_1$ and $\delta \Psi_2$ are given by Eqs.(\ref{deltapsi1-1})-(\ref{deltapsi2-1}).

From these expressions, we find that relative to the waveforms in GR, the modifications of GWs are quantified by the terms $\delta h_i$ and $\delta \Psi_i$ ($i=1,2$). In the specific case with $\delta h_i=\delta\Psi_i=0$, the formula in (\ref{final-hf}) returns to that in GR. We observe that when $\delta h_1=\delta \Psi_1=0$, it returns to the theory without parity violation. The correction term $\delta h_2$, caused by the parameter $\bar{\nu}$ in Eq.(\ref{A8}), influences only on the amplitude of $\tilde{h}(f)$. In particular, if $\delta h_2$ is frequency-independent, i.e. ${\beta_{\bar\nu}}=0$, the effect of GW amplitude modification is equivalent to modify the effective luminosity distance of GW sources \cite{nonlocal}. On the other hand, the correction term $\delta \Psi_2$, caused by the parameter $\bar{\mu}$ in Eq.(\ref{A8}), influences only on the phase of $\tilde{h}(f)$. There are consistent with the results in the previous works \cite{arai-123,ezquiaga,will}, where the specific theories in this case are discussed.

The modifications of GW waveforms, due to the PV terms in the theories of gravity, are represented by the terms $\delta h_1$ and $\delta \Psi_1$. As discussions in the Sec. \ref{sec4}, $\delta h_1$ is the result of amplitude birefringence effect between left-hand and right-hand polarization modes, which is caused by the parameter $\nu_{\rm A}$ in Eq.(\ref{A8}), and $\delta \Psi_1$ is result of amplitude birefringence effect between two circular polarization modes, caused by the parameter $\nu_{\rm A}$ in Eq.(\ref{A8}). Since in the parity-violating gravities, the evolution of polarization modes $h_+$ and $h_{\times}$ are not independent, the mixture of two modes are inevitable. For this reason, we find that both terms $\delta h_1$ and $\delta\Psi_1$ appear in the phase and amplitude modifications of $\tilde{h}(f)$. As the extension of our previous work \cite{qiao}, we find when $\beta_{\nu}=\beta_{\mu}=1$, the above results automatically return to those in \cite{qiao}. In particular, in the CS gravity, only $\delta \Psi_1$ is nonzero, and the formulas in Eq.(\ref{final-delta}) return to the corresponding quantities in \cite{cs4}. However, in general, both $\delta h_1$ and $\delta \Psi_1$ are nonzero in the theories of PV gravity. In the leading order, the modification $\delta\mathcal{A}$ (or $\delta\Psi$) linearly depends on them, and it is important to estimate their relative magnitudes. Let us assume the GW is emitted at the redshift $z\sim O(1)$, and approximately treat $\alpha_{\nu}$ and $\alpha_{\mu}$ as constants during the propagation of GW. In addition, we assume $\beta_{\nu}=\beta_{\mu}$, which is retained in most cases. Therefore, we find the ratio of two correction terms is $\delta\Psi_1/\delta h_1 \sim t_0 f$, where $f$ is the GW frequency and $t_0=13.8$ billion years is the cosmic age \cite{planck}. As known, $f\sim 100$ Hz for the ground-based GW detectors, and $f\sim 0.01$ Hz for the space-borne detectors. For both cases, we find $\delta\Psi_1$ is more than ten orders of magnitude larger than $\delta h_1$. So, we arrive at the conclusion: In the PV gravities with both velocity and amplitude birefringence effects, both the amplitude and phase corrections of GW waveform $\tilde{h}(f)$ mainly come from the contribution of velocity birefringence rather than that of amplitude birefringence.

\section{Conclusions \label{sec6}}
\renewcommand{\theequation}{6.\arabic{equation}} \setcounter{equation}{0}
With the discovery of GW sources by aLIGO and aVirgo collaborations, the testings of gravity in the strong gravitational fields become possible. Therefore, the studies on GWs in the alternative theories of gravity are important. In the series of our works, we focus on how to test the parity symmetry in gravity. In the theories of PV gravity, the symmetry between left-hand and right-hand circular polarization modes of GWs is broken. So, the effects of birefringence between these two modes occur during their propagation in the universe. In this article, we investigate the waveforms of GWs in the general PV gravity, and the existing models in the literature, including Chern-Simons modified gravity, ghost-free scalar-tensor gravity, symmetric teleparallel equivalence of GR theory, Ho\v{r}ava-Lifshitz gravity, are all the specific cases of this general theory. We find that, in general, both amplitude and velocity birefringence effects between the left-hand and right-hand polarization modes exist in these gravities, which exactly correspond to the amplitude and phase modifications of waveforms for each mode. For an individual circular polarization mode, the amplitude and/or phase of GW waveform, as well as the velocity of GW, are modified by two parameters: one describes the parity violation in gravity, and the other describes the modification of GR but keeps the parity symmetry. For this reason, in order to model-independently test the parity symmetry of gravity, we should investigate the difference between two circular polarizations, rather than focusing on an individual mode. Combining these two modes, we obtain the GW waveforms produced by the compact binary coalescence, and derive the correction terms relative to that in GR. We find that, if the parity symmetry is not broken, the correction term in the amplitude (phase) of $\tilde{h}(f)$ only comes from $\delta h_2$ ($\delta\Psi_2$), which is the amplitude (phase) modification in the each circular polarization mode. However, if the parity symmetry is violated in the theory, in addition to these terms, the corrections in the amplitude (phase) of $\tilde{h}(f)$ also comes from both $\delta h_1$  and $\delta\Psi_1$ terms, where $\delta h_1$ encodes the effect of amplitude birefringence, and $\delta\Psi_1$ represents the effect of velocity birefringence. The mixture of them in the waveform of $\tilde{h}(f)$ is caused by the fact that in the PV gravities, the general plus and cross polarization modes are not independent in their propagations. Comparing the contributions of $\delta h_1$  and $\delta\Psi_1$ in the modification of GW waveforms, we find the latter one is completely dominant, unless the velocity birefringence does not exist in the theory. All these conclusions are consistent with those derived from the previous work \cite{qiao}. We should mention that, considering the current and potential observations of ground-based and space-borne GW detectors, the explicit waveforms of GWs derived in this article can be used as the template to constrain the these gravities with parity violation. The comprehensive analysis on this topic will be carried on in a separate paper of this series of works.

\section*{Acknowledgements}
We appreciate the helpful discussions with Kai Lin, Xian Gao, Linqing Wen, Yuxiao Liu, Yi-fu Cai and Aindriu Conroy.
W.Z. is supported by NSFC Grants No. 11773028, No. 11633001, No. 11653002, No. 11421303, No. 11903030, the Fundamental Research Funds for the Central Universities, and the Strategic Priority Research Program of the Chinese Academy of Sciences Grant No. XDB23010200. J.Q., T.Z., and A.W.  are supported in part by National Natural Science Foundation of China with the Grants No. 11675143, No. 11975203, No. 11675145, and the Fundamental Research Funds for the Provincial Universities of Zhejiang in China with Grant No. RF-A2019015.


\begin{thebibliography}{399}




\bibitem{pn}
B. S. Sathyaprakash and B. F. Schutz, Living Rev.
Relativity {\bf 12}, 2 (2009).

\bibitem{spa}
M. Maggiore, {\emph{Theory and Experiments, Gravitational Waves Vol. 1}} (Oxford University Press, Oxford, England, 2007).

\bibitem{maggiore2}
M. Maggiore, {\emph{Astrophysics and Cosmology, Gravitational Waves, Volume 2. }} (Oxford University Press, Oxford, England, 2018).

\bibitem{yunes-nature}
M. C. Miller and N. Yunes, Nature {\bf 568}, 496 (2019).

\bibitem{sathya-white}
B. S. Sathyaprakash et al., {\emph{Extreme Gravity and Fundamenal Physics}}, arXiv:1903.09221.

\bibitem{gw150914}
B. P. Abbottet al.(LIGO Scientific and Virgo Collabora-tions), Phys. Rev. Lett. {\bf 116}, 061102 (2016); Phys. Rev. D {\bf 93}, 122003 (2016); Phys. Rev. Lett. {\bf 116}, 241102 (2016);116, 131103 (2016).

\bibitem{gw170817}
B. P.  Abbottet  al.(LIGO  Scientific  and  VirgoCollaborations),Phys. Rev. Lett. {\bf 119}, 161101 (2017).

\bibitem{gw-other}
B. P. Abbottet al.(LIGO Scientific and Virgo Collabora-tions), Phys. Rev. Lett. {\bf 116}, 241103 (2016); Phys. Rev. Lett. {\bf 118}, 221101 (2017); Astrophys. J. {\bf 851}, L35 (2017); Phys. Rev. Lett. {\bf 119}, 141101 (2017); arXiv:1811.12907.

\bibitem{gw150914-testGR}
B. P. Abbott et al. (LIGO Scientific and Virgo Collaborations), Phys. Rev. Lett. {\bf 116}, 221101 (2016).

\bibitem{gw170817-testGR}
B. P. Abbott et al. (LIGO Scientific and Virgo Collaborations), arXiv:1811.00364; arXiv:1903.04467.

\bibitem{gw170817-speed}
LIGO Scientific Collaboration, Virgo Collaboration, Fermi Gamma-Ray Burst Monitor, INTEGRAL, Astrophys. J. Lett., {\bf 848}, L13 (2017).



\bibitem{lorentz2}
P. Creminelli and F. Vernizzi, Phys. Rev. Lett. {\bf 119}, 251302 (2017);
J. M. Ezquiaga and M. Zumalacarregui, Phys. Rev. Lett. {\bf 119}, 251304 (2017);
L. Amendola, M. Kunz, I. D. Saltas and I. Sawicki, Phys. Rev. Lett. {\bf 120}, 131101 (2018);
E. J. Copeland, M. Kopp, A. Padilla, P. M. Saffin and C. Skordis, Phys. Rev. Lett. {\bf 122}, 061301 (2019).

\bibitem{lrr}
 M. Ishak, Living Rev. Rel. 22, 1 (2019).

\bibitem{test1}
C. M. Will, {\em Theory and experiment in gravitational physics}, (Cambridge University Press, Cambridge, 1993).

\bibitem{test2}
C. M. Will, Living Rev. Relativ. {\bf 17}, 4 (2014).

\bibitem{test3}
T. Clifton, P. G. Ferreira, A. Padilla, and C. Skordis, Phys. Rep. {\bf 513}, 1 (2012).

\bibitem{test4}
I. H. Stairs, Living Rev. Relativ. {\bf 6}, 5 (2003).


\bibitem{test5}
E. Berti, K. Yagi and N. Yunes, Gen. Relativ. Gravit. {\bf 50}, 46 (2018).

\bibitem{test6}
E. Berti, K. Yagi, H. Yang and N. Yunes, Gen. Relativ. Gravit. {\bf 50}, 49 (2018).

\bibitem{Lee-Yang}
T. D. Lee and C. N. Yang, Phys. Rev. {\bf 104}, 254 (1956).

\bibitem{CS-review}
S. Alexander and and N. Yunes, Phys. Rep. {\bf 07}, 002 (2009).



\bibitem{string}
B. A. Campbell, M. J. Duncan, N. Kalopar and K. A. Olive, Nucl. Phys. B {\bf351}, 778 (1991);
B. A. Campbell, N. Kaloper, R. Madden and K. A. Olive, Nucl. Phys. B {\bf399}, 137 (1993).




\bibitem{JackiwPi}
R. Jackiw and S. Y. Pi, Phys. Rev. D {\bf 68}, 104012 (2003).

\bibitem{cs1}
N. Yunes, R. O'Shaughnessy, B. J. Owen and S. Alexander, Phys. Rev. D {\bf 82}, 064017 (2010).

\bibitem{cs2}
K. Yagi, N. Yunes and T. Tanaka, Phys. Rev. Lett. {\bf 86}, 044037 (2012).

\bibitem{cs3}
S. H. Alexander and N. Yunes, Phys. Rev. D {\bf 97}, 064033 (2018).

\bibitem{cs4}
K. Yagi and H. Yang, Phys. Rev. D {\bf 97}, 104018 (2018).





\bibitem{ghost1}
M. Crisostomi, K. Noui, C. Charmousis and D. Langlois,
Phys. Rev. D {\bf 97}, 044034 (2018).

\bibitem{ghost2}
A. Nishizawa and T. Kobayashi, Phys. Rev. D {\bf 98}, 124018 (2018).



\bibitem{gao}
X. Gao and X. Y. Hong, arXiv:1906.07131.



\bibitem{tele}
A. Conroy and T. Koivisto, arXiv:1908.04313.

\bibitem{Horava:2009uw}
  P.~Ho\v{r}ava,
  {Phys.\ Rev.\ D {\bf 79}, 084008 (2009)}.

\bibitem{soda}
T. Takahashi and J. Soda, Phys. Rev. Lett. {\bf 102}, 231301 (2009).


\bibitem{soda2}
D.~Yoshida and J.~Soda,
  Int.\ J.\ Mod.\ Phys.\ D {\bf 27}, 1850096 (2018);
M.~Satoh and J.~Soda,
  JCAP {\bf 0809}, 019 (2008);
M.~Satoh, S.~Kanno and J.~Soda,
  Phys.\ Rev.\ D {\bf 77}, 023526 (2008);
  JHEP {\bf 1108}, 067 (2011).

\bibitem{Zhu:2011xe}
  T.~Zhu, Q.~Wu, A.~Wang and F.~W.~Shu, {Phys.\ Rev.\ D {\bf 84}, 101502 (2011)}.



\bibitem{wang_polarizing_2012}
A. Wang, Q. Wu, W. Zhao, and T. Zhu, {Phys. Rev. D {\bf 87}, 103512 (2013)}.

\bibitem{zhu_effects_2013}
T. Zhu, W. Zhao, Y. Huang, A. Wang, and Q. Wu, {Phys. Rev. D {\bf 88}, 063508 (2013)}.

\bibitem{Wang:2017brl}
A.~Wang, {Int.\ J.\ Mod.\ Phys.\ D {\bf 26}, 1730014 (2017)}.


\bibitem{qiao}
J. Qiao, T. Zhu, W. Zhao and A. Z. Wang, arXiv:1909.03815.

\bibitem{zhao2019}
W. Zhao, T. Liu, L. Q. Wen, T. Zhu, A. Z. Wang, Q. Hu and C. Zhou, arXiv:1909.13007.

\bibitem{lue}
  A.~Lue, L.~M.~Wang and M.~Kamionkowski, {Phys.\ Rev.\ Lett.\  {\bf 83}, 1506 (1999)}.

\bibitem{Alexander:2004wk} S.~Alexander and J.~Martin, {Phys.\ Rev.\ D {\bf 71}, 063526 (2005)}.

\bibitem{smolin}
C. R. Contaldi, J. Magueijo and L. Smolin, Phys. Rev. Lett. {\bf 101}, 141101 (2008).

\bibitem{nohope}
M. Gerbino, A. Gruppuso, P. Natoli, M. Shiraishi and A.
Melchiorri, JCAP {\bf 07}, 044 (2016).

\bibitem{grishchuk}
L. P. Grishchuk, Sov. Phys. JETP {\bf 40}, 409 (1975); Ann. N.Y. Acad. Sci. {\bf 302}, 439 (1977); JETP Lett. {\bf 23}, 293 (1976).



\bibitem{weinberg}
S. Weinberg, {\em  Cosmology}, (Oxford University Press, Oxford, 2008).


\bibitem{mtw}
C. W. Misner, K. S. Throne and J. A. Wheeler, {\em Gravitation}, (W. H. Freeman and Company, New York, 1973).




\bibitem{effective-field}
P. Creminelli, J. Gleyzes, J. Norena and F. Vernizzi, Phys. Rev. Lett. {\bf 113}, 231301 (2014).

\bibitem{horn}
G. W. Horndeski, Int. J. Theor. Phys.
{\bf 10}, 363 (1974); C. Deffayet, X. Gao, D. A. Steer and G. Zahariade,
Phys. Rev. D {\bf 84}, 064039 (2011);
T. Kobayashi, M. Yamaguchi and J. Yokoyama, Prog. Theor. Phys. {\bf 126},
511 (2011).

\bibitem{nonlocal}
E. Belgacem, Y. Dirian, S. Foffa and M. Maggiore, Phys. Rev. D {\bf 97}, 104066 (2018); Phys. Rev. D {\bf 98}, 023510 (2018);
E. Belgacem, Y. Dirian, S. Foffa, E. J. Howell, M. Maggiore and T. Regimbau, JCAP {\bf 08}, 015 (2019);  E. Belgacem, G. Calcagni, M. Crisostomi et al., JCAP {\bf 07}, 024 (2019).


\bibitem{arai-123}
A. Nishizawa, Phys. Rev. D {\bf 97}, 104037 (2018);
S. Arai and A. Nishizawa, Phys. Rev. D {\bf 97}, 104038 (2018); Phys. Rev. D {\bf 99}, 104038 (2019).

\bibitem{ezquiaga}
J. M. Ezquiaga and M. Zumalacarregui, Front. Astron. Space Sci. {\bf 5}, 44 (2018).


\bibitem{will}
C. M. Will, Phys. Rev. D {\bf 57}, 2061 (1998);
S. Mirshekari, N. Yunes and C. M. Will, Phys. Rev. D {\bf 85}, 024041 (2012).


\bibitem{massive}
S. Deser, R. Jackiw and S. Templeton, Phys. Rev. Lett. {\bf 48}, 975 (1982); Ann. Phys. {\bf 140}, 372 (1982).


\bibitem{tele18}
R. Aldrovandi and J. G. Pereira, {\emph{Teleparallel Gravity}}, volume 173. Springer,
23 Dordrecht, (2013).

\bibitem{tele19}
J. M. Nester and H. J. Yo. Chin. J. Phys., {\bf 37}, 113 (1999).

\bibitem{tele20}
J. B. Jimnez, L. Heisenberg, and T. Koivisto, Phys. Rev. D {\bf 98}, 044048 (2018).

\bibitem{zhu2012}
T. Zhu, F. W. Shu, Q. Wu and A. Z. Wang, Phys. Rev. D {\bf 85}, 044053 (2012).

\bibitem{34-37}
G. Amelino-Camelia, Phys. Lett. B {\bf 510}, 255 (2001); J. Magueijo and L. Smolin, Phys. Rev. Lett. {\bf 88},
190403 (2002); G. Amelino-Camelia, Nature {\bf 418}, 34 (2002).

\bibitem{dhost}
D. Langlois, M. Mancarella, K. Noui, and F. Vernizzi,
JCAP {\bf 05}, 033 (2017).

\bibitem{massive}
C. M. Will, {\emph{Theory and experiment in gravitational
physics}} (Cambridge University Press, Cambridge, UK,
1993).


\bibitem{38}
A. S. Sefiedgar, K. Nozari, and H. R. Sepangi, Phys.
Lett. B {\bf 696}, 119 (2011).


\bibitem{43-45}
R. Garattini and G. Mandanici, Phys. Rev. D {\bf 83}, 084021
(2011).

\bibitem{57}
E. Belgacem, Y. Dirian, S. Foffa and M. Maggiroe, JCAP {\bf 1803}, 002 (2018).


\bibitem{effective-planck-mass}
L. Amendola, I. Sawicki, M. Kunz and I. D. Saltas, JCAP {\bf 08}, 030 (2018).

\bibitem{cai}
Y. F. Cai, C. Li, E. N. Saridakis and L. Q. Xue, Phys. Rev. D {\bf 97}, 103513 (2018).

\bibitem{ppe}
N. Yunes and F. Pretorius, Phys. Rev. D {\bf 80} 122003 (2009);
N. Cornish, L. Sampson, N. Yunes and F. Pretorius, Phys. Rev. D {\bf 84} 062003 (2011);
K. Chatziioannou, N. Yunes and N. Cornish, Phys. Rev. D {\bf 86} 022004 (2012).



\bibitem{F+Fx}
P. Jaranowski, A. Krolak, and B. F. Schutz, Phys. Rev. D
{\bf 58}, 063001 (1998);
W. Zhao and L. Q. Wen, Phys. Rev. D {\bf 97}, 064031 (2018).




\bibitem{300}
K. Throne, in {\emph {300 Years of Gravitation}, Edited by S. Hawking and W. Isreal}, Cambridge University Press, (1987).




\bibitem{zhao2017}
X. Zhang, J. M. Yu, T. Liu, W. Zhao and A. Z. Wang, Phys. Rev. D {\bf 95}, 124008 (2017);
T. Liu, X. Zhang, W. Zhao, K. Lin, C. Zhang, S. J. Zhang, X. Zhao, T. Zhu and A. Z.  Wang, Phys. Rev. D {\bf 98}, 083023 (2018).

\bibitem{zhang2017}
X. Zhang, W. Zhao, H. Huang and Y. F. Cai, Phys. Rev. D {\bf 93}, 124003 (2016);
X. Zhang, T. Liu and W. Zhao, Phys. Rev. D {\bf 95}, 104027 (2017);
X. Zhang, W. Zhao, T. Liu, K. Lin, C. Zhang, S. J. Zhang, X. Zhao, T. Zhu and A. Z. Wang, ApJ {\bf 874}, 121 (2019);
X. Zhang, R. Niu and W. Zhao, Phys. Rev. D {\bf 100}, 024038 (2019).

\bibitem{extra}
T. P. Sotiriou, Phys. Rev. Lett. {\bf 120}, 041104 (2018).


\bibitem{planck}
Planck Collaboration, Astron. Astrophys. {\bf 571}, A31 (2014).

\bibitem{smith}
T. L. Smith, A. L. Erickcek, R. R. Caldwell, and M. Kamionkowski, Phys. Rev. D {\bf 77}, 024015 (2008).

\bibitem{yunes}
N. Yunes and D. N. Spergel, Phys. Rev. D {\bf 80}, 042004 (2009);  Y. Ali-Haimoud, Phys. Rev. D {\bf 83}, 124050 (2011).






\end{thebibliography}
\end{document}